\documentclass[conference]{IEEEtran}
\IEEEoverridecommandlockouts
\usepackage{cite}
\usepackage[hyphens]{url}
\usepackage[hidelinks]{hyperref}
\hypersetup{colorlinks=true,linkcolor=black,citecolor=black,filecolor=black,urlcolor=black,breaklinks=true}
\usepackage{float}
\usepackage{color}
\usepackage{amsmath,amssymb,amsfonts}
\usepackage{algorithmic}
\usepackage{graphicx}
\usepackage{svg}
\usepackage{textcomp}
\usepackage{xcolor}
\usepackage{multirow}
\usepackage{multicol}
\def\BibTeX{{\rm B\kern-.05em{\sc i\kern-.025em b}\kern-.08em
    T\kern-.1667em\lower.7ex\hbox{E}\kern-.125emX}}
\begin{document}

\title{Evaluation of Voltage Unbalance Metrics in Distribution Networks with High DER Penetration\\
}

%\author{\IEEEauthorblockN{Alireza Zabihi}
%\IEEEauthorblockA{\textit{ Department of Electrical Engineering} \\
%\textit{Polytechnic University of Madrid}\\
%Madrid, Spain \\
%alireza.zabihi@upm.es}
%\and
%\IEEEauthorblockN{Luis Badesa}
%\IEEEauthorblockA{\textit{ Department of Electrical Engineering} \\
%\textit{Polytechnic University of Madrid}\\
%Madrid, Spain \\
%luis.badesa@upm.es}
%\and
%\IEEEauthorblockN{Araceli Hernandez}
%\IEEEauthorblockA{\textit{ Department of Electrical Engineering} \\
%\textit{Polytechnic University of Madrid}\\
%Madrid, Spain \\
%araceli.hernandez@upm.es}
%}
\author{
\IEEEauthorblockN{Alireza Zabihi\IEEEauthorrefmark{1}, Luis Badesa\IEEEauthorrefmark{2} and Araceli Hernández\IEEEauthorrefmark{1}}
\vspace{3mm}
\IEEEauthorblockA{\IEEEauthorrefmark{1}\textit{Escuela Técnica Superior de Ingenieros Industriales (ETSII), Universidad Politecnica de Madrid (UPM), Madrid, Spain}
\\alireza.zabihi@upm.es, araceli.hernandez@upm.es}
\vspace{1mm}
\IEEEauthorblockA{\IEEEauthorrefmark{2}\textit{Escuela Técnica Superior de Ingeniería y Diseño Industrial (ETSIDI), Universidad Politecnica de Madrid (UPM), Madrid, Spain}
\\{luis.badesa@upm.es}}
}

\maketitle

\begin{abstract}
Voltage unbalance, caused by variations in voltage magnitude and phase angle, is a significant power quality issue in three-phase systems, leading to equipment inefficiencies and increased system losses. The integration of distributed energy resources (DER) into the grid adds complexity, as DER can either reduce or worsen voltage unbalance, depending on factors such as grid configuration and the distribution of loads and DER themselves. This study explores the effects of DER penetration on voltage unbalance levels and the accuracy of the different indices most commonly used to quantify this unbalance. The results highlight the varying impacts of DER on unbalance and index performance, emphasizing the need for effective strategies to assess voltage unbalance in modern distribution systems.
\end{abstract}

\begin{IEEEkeywords}
Power Quality, Voltage Unbalance, Distributed Energy Resources, Renewable Integration
\end{IEEEkeywords}

\section{Introduction}
\label{sec:introduction}
Voltage unbalance is a common issue in power quality that occurs when three-phase voltages in a system differ in magnitude, deviate from the ideal 120-degree phase separation, or both. It can cause significant operational challenges, including overheating, reduced efficiency of transformers and motors, malfunctioning protection equipment, and increased system losses \cite{kim2005comparison}, \cite{woolley2012statistical}. In low voltage distribution networks, unbalance is often due to uneven load distribution, structural asymmetries in system components, or abnormal conditions such as equipment failures and poor electrical connections \cite{singh2007some}. In addition, emerging technologies such as photovoltaics (PV), battery storage systems, and electric vehicles are increasingly recognized as contributors to voltage unbalance in modern power systems \cite{antic2020comprehensive}, \cite{rodriguez2021probabilistic}. DER often generate power intermittently and unevenly in phases, introducing dynamic and time-varying changes in load distribution \cite{girigoudar2019relationships}. This variability disrupts traditional methods for measuring and maintaining phase balance and increases the likelihood of unbalanced conditions.

The first step in addressing voltage unbalance is to develop a feasible, accurate, and efficient method for measuring and calculating it. Various organizations, including IEEE, IEC, NEMA and CIGRE, have proposed definitions and methodologies for calculating voltage unbalance, each with its advantages and limitations. These methods differ in terms of input measurement requirements, result accuracy, and potential numerical challenges, which are important for being well integrated into broader problems \cite{girigoudar2020impact}, \cite{churkin2024quantifying}, as well as accessibility of the necessary input data. The relationships between these methodologies have been explored in previous studies. For example, in \cite{singh2007some}, a sensitivity analysis was performed but was limited to examining a single phase with magnitude or phase deviation. Another study \cite{chen2013examination} used a genetic algorithm to identify differences among various definitions of voltage unbalance. Furthermore, in \cite{girigoudar2019relationships}, a combination of analytical approximations and numerical methods was used to establish relationships and bounds between certain definitions of voltage unbalance. Many of these studies have not thoroughly examined existing definitions, often focusing on a narrow set of voltage conditions without offering clear theoretical error bounds for approximate metrics. Regarding PV panels integration, studies have analyzed voltage unbalance in residential low-voltage networks, considering the influence of single-phase rooftop PV panels locations and capacities through sensitivity and probabilistic approaches \cite{shahnia2011voltage}. The results of stochastic evaluations reveal that increased penetration of PV panel inverters may pose the risk of unacceptable voltage unbalance \cite{schwanz2016stochastic}.

The main contributions of this paper are listed below.
\begin{itemize} 
\item To investigate variations in voltage unbalance definitions by analyzing all different definitions within a realistic distribution network, taking into account residual loads, and evaluating their performance under practical operating conditions. 
\item To evaluate the impact of PV panels integration by assessing how this distributed generation influences the accuracy and effectiveness of voltage unbalance definitions in modern distribution networks. 
\end{itemize}

The remainder of the paper is structured as follows. Section~\ref{sec:Def&Com} presents various definitions of voltage unbalance, highlighting their advantages, limitations, and trade-offs. We also briefly discuss the bounds and relationships between these definitions. Section~\ref{sec:Sim&Method} details the simulation analysis and results based on a realistic distribution network. Section~\ref{sec:PV Int} examines the impact on voltage unbalance of integrating PV panels into the network. Finally, Section~\ref{sec:Conclusion} provides the conclusion of the study.

\section{Voltage Unbalance Definitions and Comparison}
\label{sec:Def&Com}
 In a three-phase system, voltages are typically expressed as:
\begin{equation}
\begin{aligned}
\mathbf{V}_a &= V_a \angle \theta_a \\
\mathbf{V}_b &= V_b \angle \theta_b \\
\mathbf{V}_c &= V_c \angle \theta_c
\end{aligned}
\label{eq:voltage_phasors}
\end{equation}
Upcoming indexes are defined using the above notation, where phasor voltages are denoted in bold and voltage magnitudes are represented in regular font.

\subsection{IEC Definition (True Definition)}
This definition, often referred to as the `true definition', calculates the phasors of the positive and negative sequence components \cite{IEC}:
\begin{equation}
\mathbf{V_1} = \frac{\mathbf{V_a} + a \cdot \mathbf{V_b} + a^2 \cdot \mathbf{V_c}}{3}
\end{equation}
\begin{equation}
\mathbf{V_2} = \frac{\mathbf{V_a} + a^2 \cdot \mathbf{V_b} + a \cdot \mathbf{V_c}}{3}
\end{equation}

where \( a = e^{j 120^\circ} = -\frac{1}{2} + j \frac{\sqrt{3}}{2} \).

Next, it is possible to calculate the Voltage Unbalance Factor (VUF) by dividing negative and  positive magnitudes:
\begin{equation}
\text{VUF} = \frac{V_2}{V_1} \cdot 100
\label{eq:VUF}
\end{equation}

This definition requires the measurement of both voltage magnitudes and relative voltage angles, which are often unavailable. Although it serves as the original definition and a baseline for comparing other definitions, it is highly non-linear and difficult to calculate. It should be noted that IEEE standard 1159 also provides this definition \cite{IEEE1159}.
\subsection{NEMA Definition}
Calculates the average line voltage and the maximum deviation from the average \cite{LVUR}:
\begin{equation}
V_{\text{avg}}^L = \frac{V_{ab} + V_{bc} + V_{ca}}{3}
\label{eq:avg_voltage}
\end{equation}
\begin{equation}
\Delta V^L = \max\left( \left| V_{ab} - V_{\text{avg}}^L \right|, \left| V_{bc} - V_{\text{avg}}^L \right|, \left| V_{ca} - V_{\text{avg}}^L \right| \right)
\label{eq:max_deviation}
\end{equation}
Then it is possible to calculate the Line Voltage Unbalance Ratio (LVUR):
\begin{equation}
\text{LVUR} = \frac{\Delta V^L}{V_{\text{avg}}^L} \cdot 100
\end{equation}

This definition requires measuring line voltage magnitudes and is highly non-linear due to the presence of the maximum operator, making it difficult to implement in broader power system simulation routines.

\subsection{CIGRE Definition}
\label{ss:cigre_unbalance}
This definition first requires the calculation of a parameter \(\beta\) \cite{CIGRE}:
\begin{equation}
\beta = \frac{V_{ab}^4 + V_{bc}^4 + V_{ca}^4}{\left(V_{ab}^2 + V_{bc}^2 + V_{ca}^2\right)^2}
\label{eq:beta}
\end{equation}
Then it is possible to calculate the CIGRE Unbalance Factor:
\begin{equation}
\text{Unbalance Factor} = \sqrt{\frac{1 - \sqrt{3 - 6\beta}}{1 + \sqrt{3 - 6\beta}}} \cdot 100
\label{eq:cigre_unbalance}
\end{equation}
This definition only requires the measurement of line voltage magnitudes, while being highly nonlinear. On the other hand, it is an exact reformulation of (\ref{eq:VUF}), as demonstrated in \cite{singh2007some}, \cite{chen2013examination}.

\subsection{IEEE First Definition (Standard 141)}
This metric calculates the average phase voltage and the maximum deviation from the average \cite{PVUR}:
\begin{equation}
V_{\text{avg}}^P = \frac{V_a + V_b + V_c}{3}
\label{eq:phase_avg_voltage}
\end{equation}
\begin{equation}
\Delta V^P = \max\left(\left| V_a - V_{\text{avg}}^P \right|, \left| V_b - V_{\text{avg}}^P \right|, \left| V_c - V_{\text{avg}}^P \right|\right)
\label{eq:max_phase_deviation}
\end{equation}
Then, the Phase Voltage Unbalance Ratio (PVUR1) is calculated as:
\begin{equation}
\text{PVUR1} = \frac{\Delta V^P}{V_{\text{avg}}^P} \cdot 100
\label{eq:phase_unbalance_ratio}
\end{equation}
This definition requires the measurement of phase voltage magnitudes and is highly nonlinear due to the presence of the maximum operator, making it difficult to implement.
\subsection{IEEE Second Definition (Standard 112 and 936)}
This metric is obtained by calculating the average phase voltage and the difference between the highest and the lowest voltage magnitude \cite{PVUR2-112}, \cite{PVUR2-936}:
\begin{equation}
V_{\text{avg}}^P = \frac{V_a + V_b + V_c}{3}
\label{eq:phase_avg_voltage_2}
\end{equation}
\begin{equation}
V_{\text{max}} = \max\left(V_a, V_b, V_c\right)
\label{eq:phase_max_voltage}
\end{equation}
\begin{equation}
V_{\text{min}} = \min\left(V_a, V_b, V_c\right)
\label{eq:phase_min_voltage}
\end{equation}
The Phase Voltage Unbalance Ratio (PVUR2) is calculated as:
\begin{equation}
\text{PVUR2} = \frac{V_{\text{max}} - V_{\text{min}}}{V_{\text{avg}}^P} \cdot 100
\label{eq:phase_unbalance_ratio_2}
\end{equation}
This definition requires the measurement of phase voltage magnitudes due to the presence of the maximum operator, so it faces similar challenges as previous indices.

\subsection{Comparison}
This study investigates a typical unbalanced condition in a realistic three-phase system, characterized by the following typical ranges for voltage magnitudes and phase angle deviations:
\begin{equation}
0.94 \leq \text{Voltage Magnitudes (p.u.)} \leq 1.1
\label{eq:voltage_magnitude_range}
\end{equation}
\begin{equation}
-5^\circ \leq \text{Phase Angle Deviation (degrees)} \leq 5^\circ
\label{eq:phase_angle_deviation_range}
\end{equation}

The objective is to determine the lower and upper limits of each voltage unbalance index corresponding to a specific Voltage Unbalance Factor (VUF) value: 
\begin{equation}
\text{VU Ind. }_{\text{min}} \leq \text{VUF} \leq \text{VU Ind. }_{\text{max}}
\label{eq:vu_index_range}
\end{equation}
\begin{equation}
\text{VU Ind. }_{\text{min}} / \text{VUF} \leq 1 \leq \text{VU Ind. }_{\text{max}} / \text{VUF}
\label{eq:normalized_vu_index_range}
\end{equation}
\begin{equation}
\text{Lower Bound} \leq 1 \leq \text{Upper Bound}
\label{eq:general_bound}
\end{equation}

\begin{table}[!t]
\renewcommand{\arraystretch}{1.3}
\caption{Relative index bounds for different voltage unbalance metrics when 1\% $<$ VUF $<$ 2\%}
\begin{center}
\begin{tabular}{|c|c|c|}
\hline
Relative Indexes & Upper Bound & Lower Bound \\
\hline
LVUR/VUF & 1.005 & 0.866 \\
\hline
CIGRE/VUF & 1.000 & 1.000 \\
\hline
PVUR1/VUF & 10.728 & 0.000 \\
\hline
PVUR2/VUF & 16.092 & 0.000 \\
\hline
\end{tabular}
\label{tab:relative_indexes1}
\end{center}
\end{table}

\begin{table}[!t]
\renewcommand{\arraystretch}{1.3}
\caption{Relative index bounds for different voltage unbalance metrics when 2\% $<$ VUF $<$ 3\%}
\begin{center}
\begin{tabular}{|c|c|c|}
\hline
Relative Indexes & Upper Bound & Lower Bound \\
\hline
LVUR/VUF & 1.007 & 0.866 \\
\hline
CIGRE/VUF & 1.000 & 1.000 \\
\hline
PVUR1/VUF & 5.360 & 0.000 \\
\hline
PVUR2/VUF & 8.040 & 0.000 \\
\hline
\end{tabular}
\label{tab:relative_indexes3}
\end{center}
\end{table}

Numerical results for VUF values within the range of 1\% to 2\% are presented in Table~\ref{tab:relative_indexes1}, whereas those corresponding to the 2\% to 3\% range are shown in Table~\ref{tab:relative_indexes3}.
These results align closely with the conclusions of previous studies \cite{chen2013examination}, \cite{girigoudar2019relationships}, further validating the observed relationships among voltage unbalance metrics.

\section{Simulation and Methodology}
\label{sec:Sim&Method}

\begin{figure}[!t]
\centering
\includegraphics[width=3.3in]{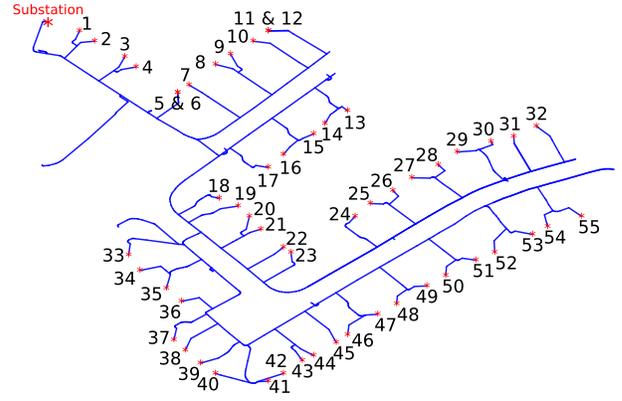}
\caption{Load buses of the IEEE European Low Voltage Test Feeder.}
\label{fig1}
\end{figure}

In order to evaluate the variations of the indexes introduced in Section \ref{sec:Def&Com} in a distribution network under realistic conditions, the IEEE European LV test feeder \cite{ieee2015} is adopted. This grid includes 55 load buses, with a nominal line voltage of 0.416 kV. The single-line diagram of this grid, indicating the load buses, is shown in Fig.~\ref{fig1}.

The results of a power flow analysis, carried out through OpenDSS \cite{opendss} were investigated for three different scenarios with almost the same amount of total load without violating voltage constraints:

\begin{itemize} 
\item Scenario I: Low voltage unbalance scenario with approximately 169 kW of single phase loads distributed as 31.7\%, 39.5\%, and 28.8\% in phases a, b, and c, respectively. 
\item Scenario II: Medium voltage unbalance level with approximately 171 kW of single phase loads distributed as 22.2\%, 31.5\%, and 45.3\% in phases a, b, and c, respectively. 
\item Scenario III: High voltage unbalance level with approximately 153 kW of single phase loads distributed as 22.1\%, 59.3\%, and 18.6\% in phases a, b, and c, respectively. 
\end{itemize}

\begin{figure}[!t]
\centering
\includegraphics[width=3.3in]{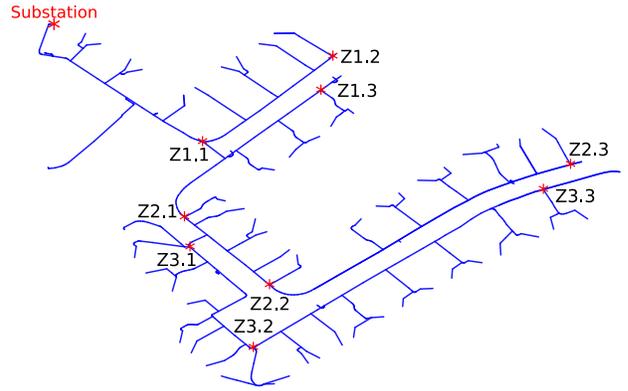}
\caption{Buses representing three main distribution zones of the IEEE European low voltage test feeder.}
\label{fig2}
\end{figure}

The voltage unbalance indices are measured in the buses of the main distribution line, which is divided into three primary distribution zones. 
To represent the overall situation of each zone, three buses are selected in each: one at the beginning, one at the middle, and one at the end of the lines. These representative buses are illustrated in Fig.~\ref{fig2}.

\begin{figure}[!t]
\centering
\includegraphics[width=3.2in]{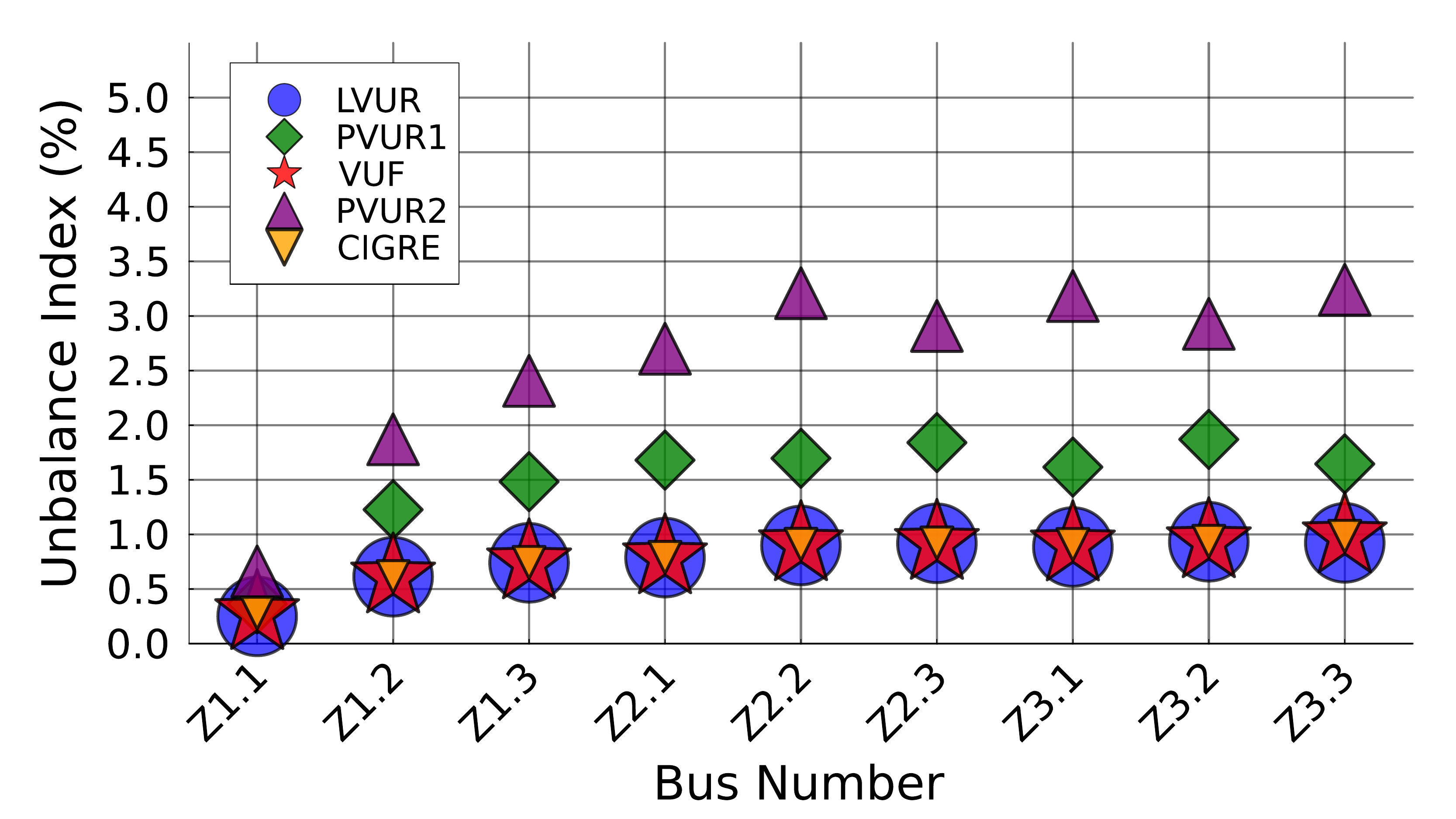}
\caption{Different voltage unbalance indexes in scenario I.}
\label{fig3.1}
\end{figure}
\begin{figure}[!t]
\centering
\includegraphics[width=3.2in]{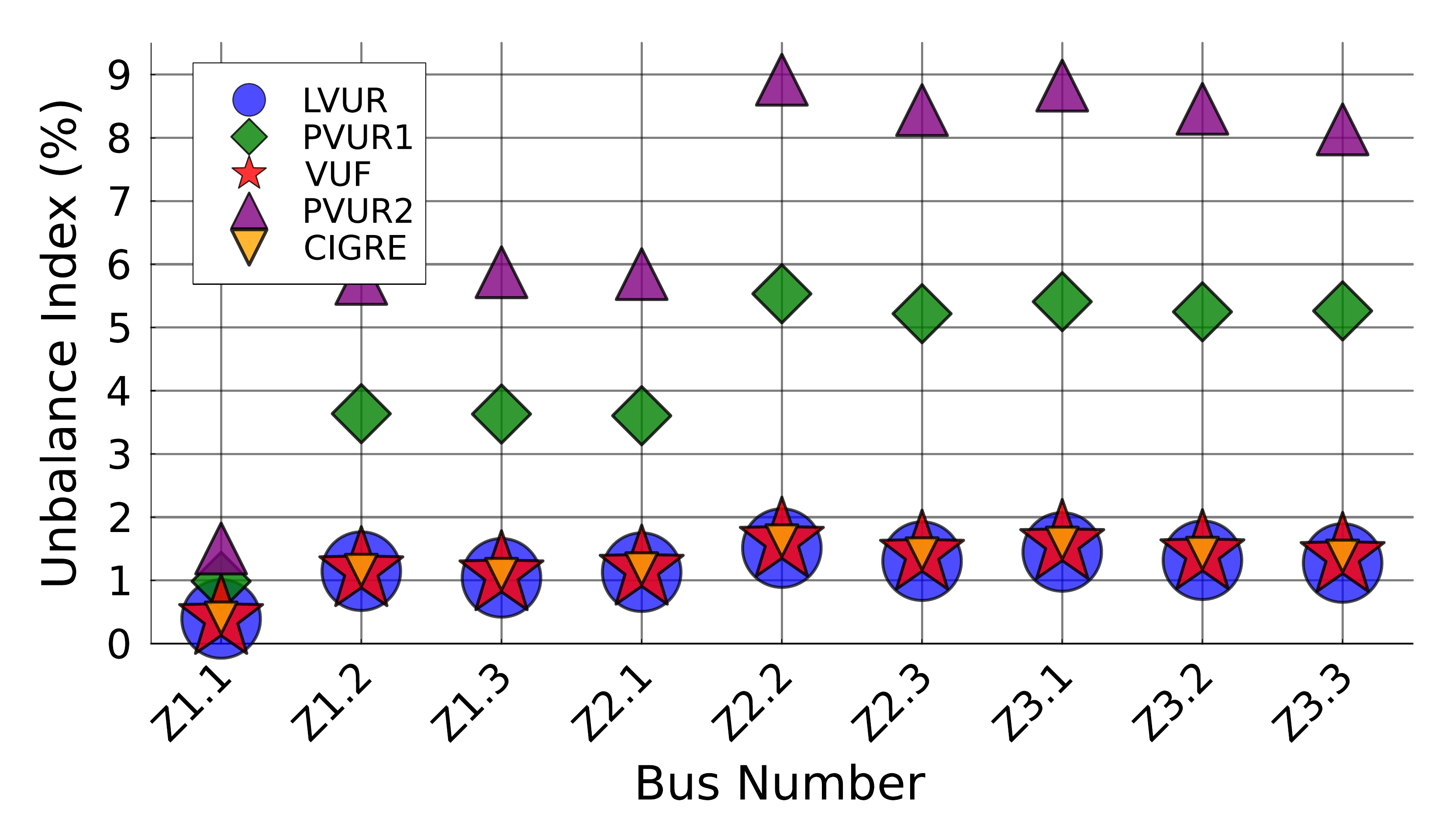}
\caption{Different voltage unbalance indexes in scenario II.}
\label{fig3.2}
\end{figure}
\begin{figure}[!t]
\centering
\includegraphics[width=3.2in]{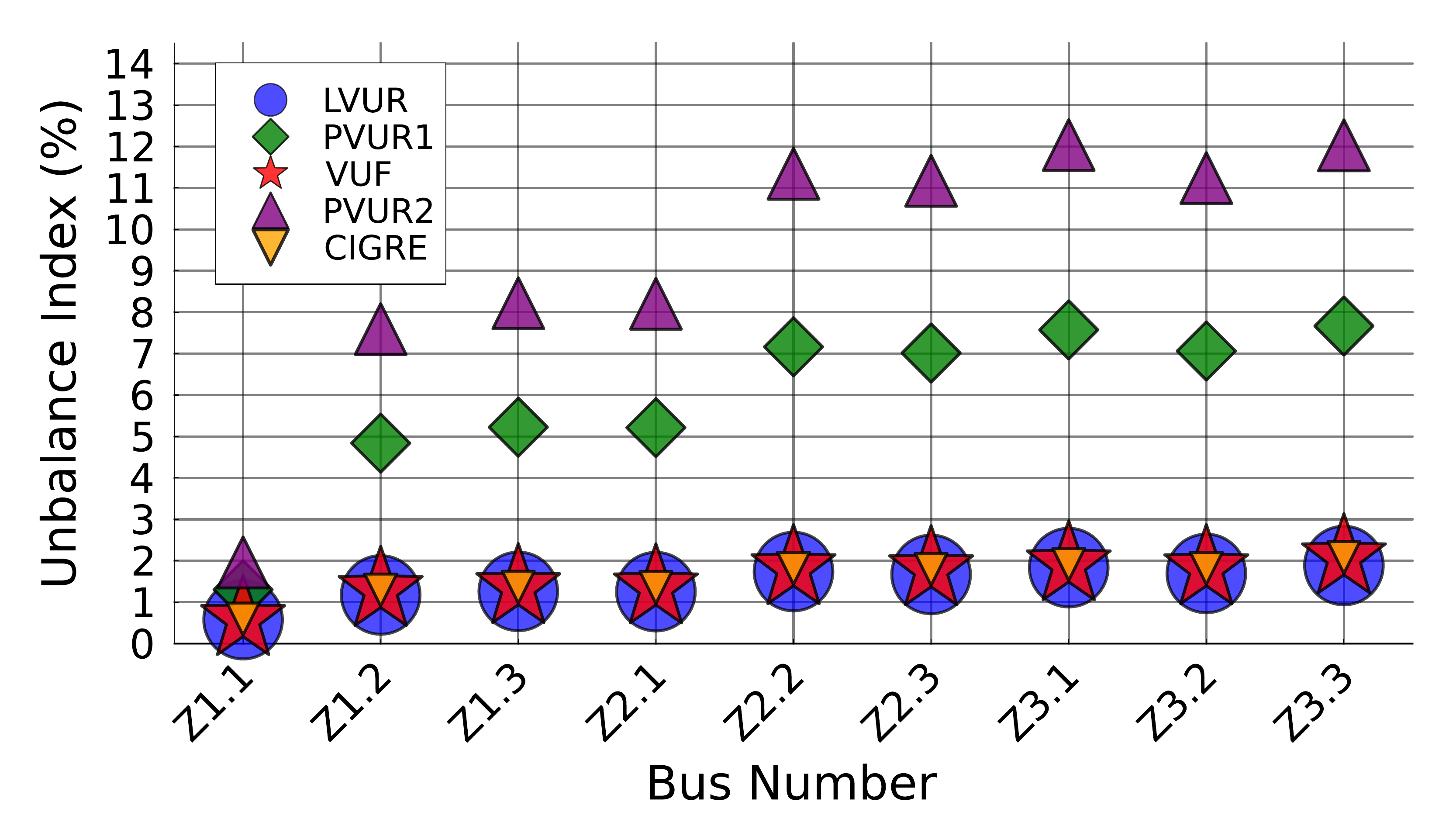}
\caption{Different voltage unbalance indexes in scenario III.}
\label{fig3.3}
\end{figure}

The results of the calculation of voltage unbalance with different definitions are presented in Fig.~\ref{fig3.1} for Scenario I, Fig.~\ref{fig3.2} for Scenario II and Fig.~\ref{fig3.3} for Scenario III.

As illustrated, the accuracy of LVUR relative to the `True definition' shows it to be a reasonable estimation of the VUF, as line voltage measurements inherently account for both voltage magnitudes and phase angle deviations. However, LVUR may underestimate VUF in certain cases, as highlighted in Table \ref{tab:relative_indexes1} and Table \ref{tab:relative_indexes3}. Additionally, when a zero-sequence component exists, line voltage measurements cannot capture it, introducing a bias in the measurements. This limitation is discussed in detail in \cite{chen2013examination}. 

The CIGRE index, on the other hand, is an exact reformulation of the VUF, as described in Section \ref{ss:cigre_unbalance}. Therefore, this index has exactly the same results as the IEC definition.

PVUR1 and PVUR2 exhibit lower accuracy, as they do not account for phase-angle deviations. Consequently, in buses where phase angle deviations are significant, these indices deviate considerably from the VUF value.

These inaccuracies can be categorized by different input data and then different formulation even with the same set of inputs. These comparisons have been made without integration of PV panels; their impact will be investigated next.

\section{Integration of PV Panels}
\label{sec:PV Int}
 To analyze the impact of integrating PV panels into residual grids, 40 single-phase PV panels, each with a nominal constant output power of 2.5 kW and a nominal output phase voltage of 240 V \cite{rodriguez2021probabilistic}, are incorporated into the network. These PV generators are assumed to be of grid-following type, as these are the inverters commonly used in low-voltage distribution networks \cite{li2022revisiting}. 

\begin{figure}[!t]
\centering
\includegraphics[width=3.2in]{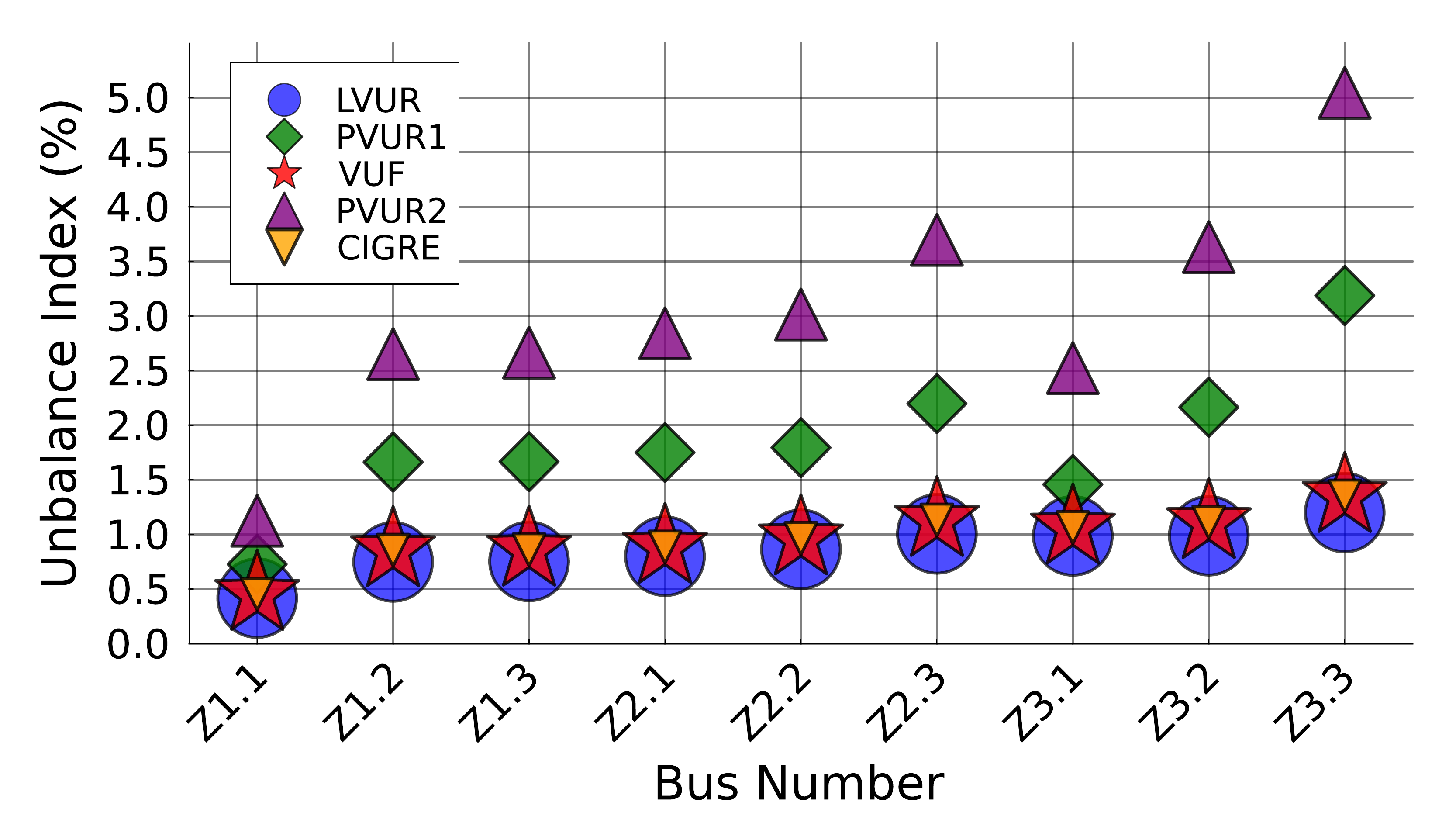}
\caption{Different voltage unbalance indexes in scenario I after PV integration.}
\label{fig3.10}
\end{figure}

\begin{figure}[!t]
\centering
\includegraphics[width=3.2in]{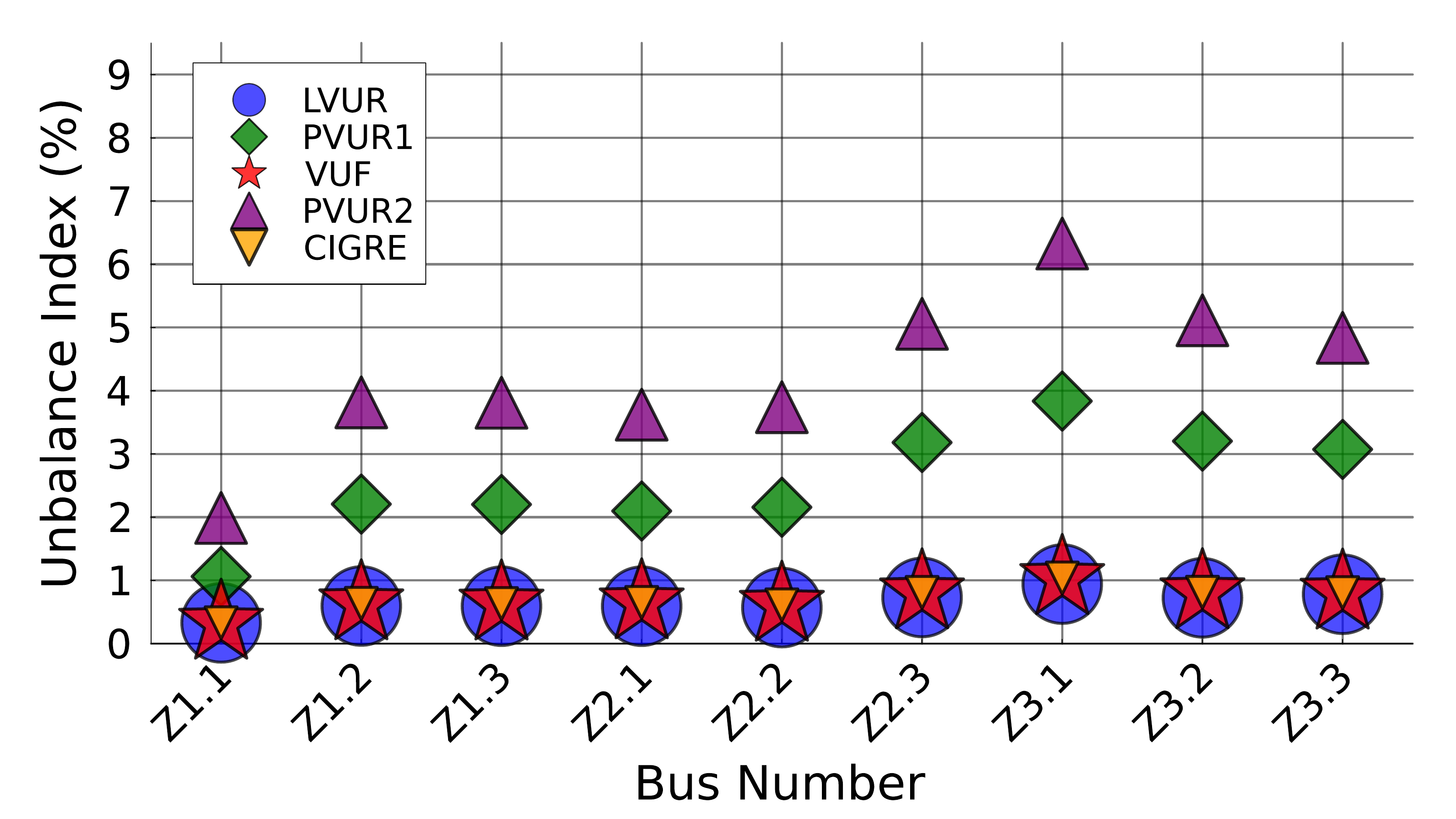}
\caption{Different voltage unbalance indexes in scenario II after PV integration.}
\label{fig3.20}
\end{figure}

\begin{figure}[!t]
\centering
\includegraphics[width=3.2in]{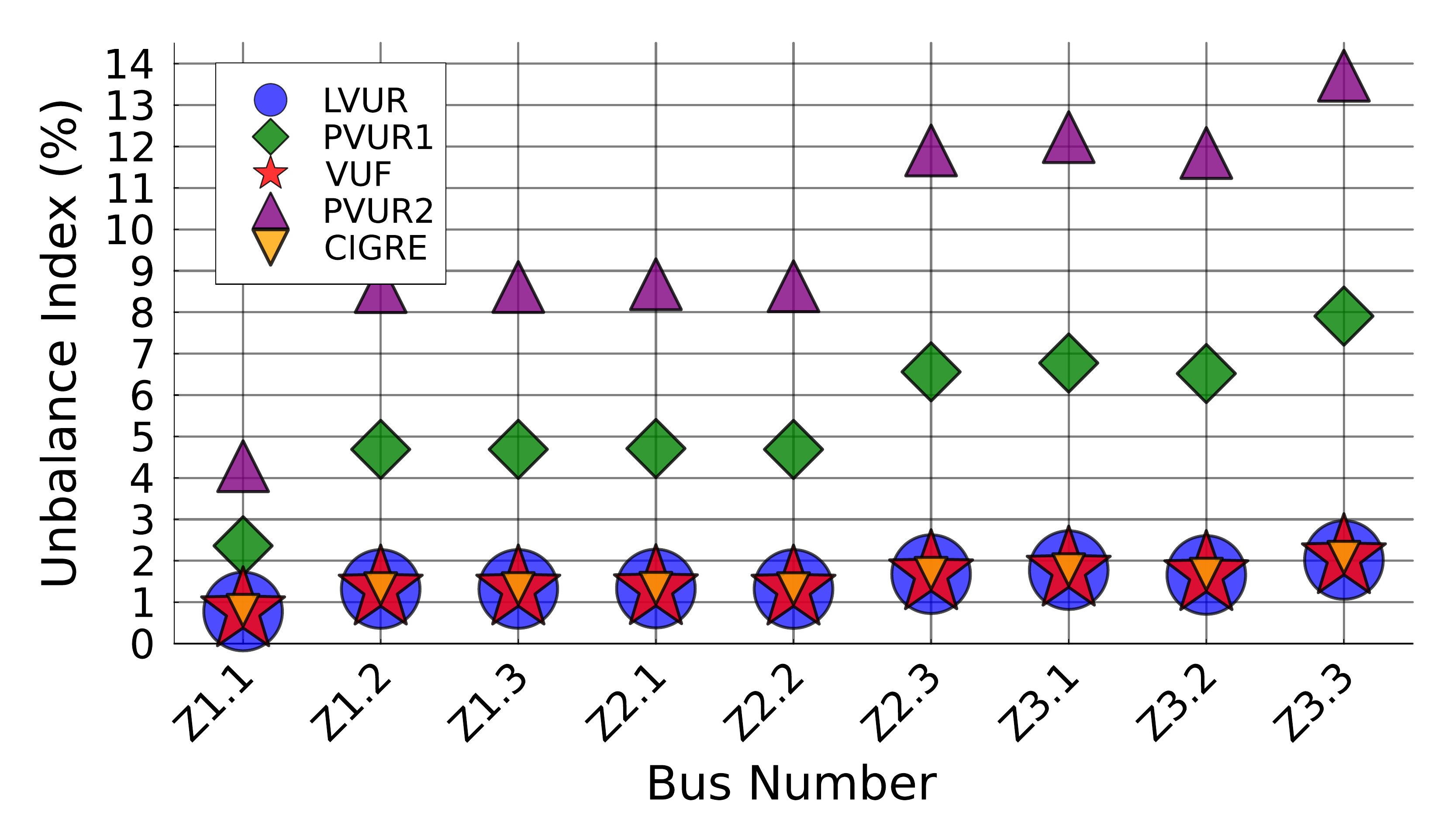}
\caption{Different voltage unbalance indexes in scenario III after PV integration.}
\label{fig3.30}
\end{figure}

The results and changes in the voltage unbalance level for the same scenarios considered in the previous section, but now with PV generation present, are illustrated in Figs.~\ref{fig3.10}, \ref{fig3.20}, and~\ref{fig3.30}. It should be noted that the buses to which the PV panels are connected are the same in all three scenarios, however, the single-phase PV generators are not distributed across phases following the exact same percentages as those used for loads, which were described in Section~\ref{sec:Sim&Method}.

\begin{table}[!t]
\renewcommand{\arraystretch}{1.3}
\centering
\caption{VUF statistics before and after integration of DER}
\begin{tabular}{|c|c|c|c|}
\hline
\textbf{VUF Stats} & \textbf{Scenario I} & \textbf{Scenario II} & \textbf{Scenario III} \\ \hline
\multicolumn{4}{|c|}{\textbf{Before PV Integration}} \\ \hline
\textbf{Max}       & 0.982              & 1.625               & 2.081                \\ \hline
\textbf{Min}       & 0.279              & 0.404               & 0.591                \\ \hline
\textbf{Mean}      & 0.787              & 1.255               & 1.558                \\ \hline
\multicolumn{4}{|c|}{\textbf{After PV Integration}} \\ \hline
\textbf{Max}       & 1.351              & 1.036               & 2.083                \\ \hline
\textbf{Min}       & 0.454              & 0.329               & 0.802                \\ \hline
\textbf{Mean}      & 0.963              & 0.702               & 1.484                \\ \hline
\end{tabular}
\label{tab:vuf_stats}
\end{table}

The high penetration of DER reduces dependence on the central feeder because the PV panels supply the local loads. However, since DER are typically grid-following, they can either reduce or increase voltage unbalance levels, depending on the grid configuration and the allocation of loads and PV panels. The results of the changes in the voltage unbalance levels before and after PV integration are presented in Table \ref{tab:vuf_stats}. It shows that in scenario I, the unbalance increases on average, in scenario II, it decreases on average, and in scenario III, the change is minimal, with almost negligible improvement.

To investigate the impact of DER integration on the accuracy of different voltage unbalance definitions, the absolute estimation error for each voltage unbalance index can be calculated with reference to the VUF index considered as the true value, as shown below:
\begin{equation}
\text{Absolute Error} = \left| \text{VU Ind.} - \text{VUF} \right|
\label{eq:absolute_error}
\end{equation}

\begin{table}[!t]
\renewcommand{\arraystretch}{1.3}
\centering
\caption{Absolute error of indices before and after DERintegration}
\label{tab:absolute_error_indices}
\begin{tabular}{|c|c|c|c|c|}
\hline
\textbf{Absolute Error} & \multirow{2}{*}{\textbf{Stats}} & \multicolumn{3}{c|}{\textbf{Scenarios}} \\ \cline{3-5} 
\textbf{of Indices (\%)} &  & \textbf{I} & \textbf{II} & \textbf{III} \\ \hline
\multicolumn{5}{|c|}{\textbf{Before PV Integration}} \\ \hline
\multirow{3}{*}{$\left| \text{LVUR} - \text{VUF} \right|$} & \textbf{Max} & 0.059 & 0.139 & 0.196 \\ \cline{2-5}
                    & \textbf{Min} & 0.001 & 0.013 & 0.013 \\ \cline{2-5}
                    & \textbf{Mean} & 0.015 & 0.080 & 0.103 \\ \hline
\multirow{3}{*}{$\left| \text{PVUR1} - \text{VUF} \right|$} & \textbf{Max} & 0.934 & 3.908 & 5.662 \\ \cline{2-5}
                    & \textbf{Min} & 0.085 & 0.581 & 0.710 \\ \cline{2-5}
                    & \textbf{Mean} & 0.705 & 3.026 & 4.338 \\ \hline
\multirow{3}{*}{$\left| \text{PVUR2} - \text{VUF} \right|$} & \textbf{Max} & 2.300 & 7.295 & 10.125 \\ \cline{2-5}
                    & \textbf{Min} & 0.383 & 1.100 & 1.370 \\ \cline{2-5}
                    & \textbf{Mean} & 1.781 & 5.609 & 7.753 \\ \hline
\multicolumn{5}{|c|}{\textbf{After PV Integration}} \\ \hline
\multirow{3}{*}{$\left| \text{LVUR} - \text{VUF} \right|$} & \textbf{Max} & 0.151 & 0.094 & 0.064 \\ \cline{2-5}
                    & \textbf{Min} & 0.037 & 0.000 & 0.006 \\ \cline{2-5}
                    & \textbf{Mean} & 0.101 & 0.052 & 0.018 \\ \hline
\multirow{3}{*}{$\left| \text{PVUR1} - \text{VUF} \right|$} & \textbf{Max} & 1.836 & 2.800 & 5.823 \\ \cline{2-5}
                    & \textbf{Min} & 0.274 & 0.733 & 1.559 \\ \cline{2-5}
                    & \textbf{Mean} & 0.882 & 1.855 & 3.950 \\ \hline
\multirow{3}{*}{$\left| \text{PVUR2} - \text{VUF} \right|$} & \textbf{Max} & 3.692 & 5.292 & 11.629 \\ \cline{2-5}
                    & \textbf{Min} & 0.672 & 1.659 & 3.483 \\ \cline{2-5}
                    & \textbf{Mean} & 2.058 & 3.555 & 8.350 \\ \hline
\end{tabular}
\end{table}

\begin{table}[!t]
\renewcommand{\arraystretch}{1.3}
\centering
\caption{Average changes of indices (VUF is the baseline)}
\begin{tabular}{|c|c|c|c|}
\hline
\textbf{Average Changes} & \textbf{Scenario I} & \textbf{Scenario II} & \textbf{Scenario III} \\ \hline
\textbf{LVUF}                & 0.086            & -0.028            & -0.085             \\ \hline
\textbf{PVUR1}               & 0.177            & -1.171             & -0.388             \\ \hline
\textbf{PVUR2}               & 0.277           & -2.054            & 0.597               \\ \hline
\end{tabular}
\label{tab:avg_improvement_VUInd}
\end{table}

The accuracy of the voltage unbalance indices before and after DER integration is presented in Table \ref{tab:absolute_error_indices}. To assess improvement or deterioration in the accuracy of the estimations, Table \ref{tab:avg_improvement_VUInd} provides the changes in accuracy on average. It is important to note that, as accuracy is evaluated based on absolute error, negative values indicate improvement (reduced error), while positive values indicate deterioration (increased error). 
In Scenario I, the VUF experiences a slight increase, accompanied by a minor decrease in accuracy. In Scenario II, the VUF shows a significant decrease, while the accuracy exhibits a substantial increase. In Scenario III, both the VUF and accuracy undergo minimal changes.

In conclusion, when the voltage unbalance level, defined by the true VUF, improves after PV integration, the accuracy of various indices also improves because of reductions in both voltage magnitude and phase angle deviations. This improvement is particularly significant in phase voltage-based indices (PVUR1 and PVUR2) since these indices do not account for phase angle deviations, and thus their accuracy benefits greatly from the reduction of this factor. In contrast, when the level of voltage imbalance deteriorates after PV integration, the accuracy of the indices also decreases, following a similar logical explanation.

\section{Conclusion}
\label{sec:Conclusion}
This study evaluated the accuracy and applicability of various voltage unbalance indices under realistic network conditions. LVUR was found to be a reasonable approximation of the true definition (VUF), because line voltage measurements inherently consider both voltage magnitudes and phase angle deviations. However, LVUR may underestimate VUF in certain scenarios. The CIGRE index, being an exact reformulation of VUF, offers superior accuracy, while PVUR1 and PVUR2 were shown to be less reliable due to their exclusion of phase angle deviations, resulting in significant discrepancies when phase angle deviations are prominent.

The high penetration integration of DER impacts voltage unbalance and index accuracy depending on the allocation of DER and loads and network configuration. When PV integration reduces VUF, all indices improve, particularly those based on phase voltage (PVUR1, PVUR2), which benefit from decreased phase angle deviations. Conversely, when the VUF increases, the index accuracy deteriorates, following the same dependency on the voltage magnitude and phase angle deviations.

In conclusion, this work highlights the interplay between voltage unbalance definitions and the evolving characteristics of modern distribution networks with high DER penetration. Understanding these relationships is critical for maintaining power quality and guiding research aiming to study broader problems related to voltage unbalance and the development of mitigation strategies.

Our ongoing work examines various factors that influence voltage imbalance in distribution grids and considers the interplay between computational efficiency and the accuracy of related metrics to address broader challenges.

\section{Acknowledgement}
This work was supported by MICIU/AEI/10.13039/501100011033 and ERDF/EU under grant PID2022-141609OB-I00, %under grants PID2022-141609OB-I00 and PID2023-150401OA-C22.
and by the Madrid Government (Comunidad de Madrid-Spain) under the Multiannual Agreement 2023-2026 with Universidad Politécnica de Madrid, ``Line A - Emerging PIs''.
The work of Alireza Zabihi was supported by the 2023 FPI-UPM call for Predoctoral Contracts within the framework of the 2021-2023 State Plan for Scientific, Technical, and Innovative Research.

\IEEEtriggeratref{5}

%\bibliographystyle{IEEEtran}
%\bibliography{mybib.bib}
\bibliographystyle{IEEEtran} 
\bibliography{Bibliography}
\end{document}